\documentclass[pre,twocolumn,preprintnumbers,amsmath,amssymb]{revtex4}

\usepackage{amsfonts}
\usepackage{bbold}
\usepackage{mathrsfs}
\usepackage{color}
\usepackage{epstopdf}

\usepackage{graphicx}
\usepackage{bm}

\usepackage{amssymb, dsfont}

\newcommand{\nn}{\nonumber }

\newcommand{\rr}{{\mathbf r}}

\newcommand{\p}{\partial}

\newcommand{\BEQ}{\begin{equation}}
\newcommand{\EEQ}{\end{equation}}
\newcommand{\BEA}{\begin{eqnarray}}
\newcommand{\EEA}{\end{eqnarray}}

\newcommand{\rv}{\mathbf{r}}

\newcommand{\Fv}{\mathbf{F}}
\newcommand{\fs}{\mathbf{f_s}}

\newcommand{\ee}{\mathbf{e}}
\newcommand{\vv}{\mathbf{v}}

\begin{document}

\title{Flocking Transition in Confluent Tissues}

\author{Fabio Giavazzi$^1$}
\author{Matteo Paoluzzi$^2$}
\author{Marta Macchi$^1$}
\author{Dapeng Bi$^3$}
\author{Giorgio Scita$^{4,5}$}
\author{Lisa Manning$^2$}
\author{Roberto Cerbino$^1$}
\author{Cristina Marchetti$^2$}

\affiliation{
$^1$  Universit\`a degli Studi di Milano, Dipartimento di Biotecnologie Mediche e Medicina Tradizionale. \\
$^2$  Physics Department and Soft Matter Program, Syracuse University, Syracuse, NY 13244, USA. \\
$^3$ Department of Physics, Northeastern University, Boston, MA. \\
$^4$ IFOM-FIRC Institute of Molecular Oncology, Milan, Italy. \\
$^5$ Universit\`a degli Studi di Milano, Dipartimento di Oncologia e Emato-Oncologia.
}


\begin{abstract}
Collective cell migration underlies important biological processes, such as embryonic development, wound healing and cancer invasion. While many aspects of single cell movements are now well established, the mechanisms leading to displacements of cohesive cell groups are still poorly understood. 
To elucidate the emergence of collective migration in mechanosensitive cells, we examine a self-propelled Voronoi (SPV) model of confluent tissues with an orientational
feedback that aligns a cell's polarization with its local migration velocity. While shape and motility are known to regulate a density-independent liquid-solid transition in tissues, we find that aligning interactions facilitate collective motion and promote solidification. Our model reproduces the behavior observed in jammed epithelial monolayers, which are unjammed by the addition of the endocytic protein RAB5A that promotes cell motility by inducing large scale coherent migratory patterns and local fluidization. 
\end{abstract}


\maketitle


\section*{Introduction}
The main cause of mortality in cancer patients is the spreading of primary cancer cells that generate metastatic foci through complex and still poorly understood processes. The invasiveness of metastatic cells is
facilitated by  their plasticity, i.e., their ability to adapt to the microenvironment and change their identity to invade healthy tissues and proliferate \cite{Marjanovic13,Nieto13}. Key to invasion is cell migration. Migratory phenotypes are intrinsically flexible and include both single and collective cell motility modes \cite{Clark15,Haeger15}. For example, migrating cells can display both mesenchymal and epithelial phenotypes or frequently interconvert between these two states in a process commonly referred to as Epithelial-to-Mesenchymal Transition (EMT). Cells undergoing EMT detach from the surrounding cells and become hyper-motile to revert their state back to epithelial to seed distant metastatic foci \cite{Thompson05}. Although the role of EMT in cancer formation is still debated \cite{Ledford11}, several studies pointed out the importance of EMT in metastasis formation and cancer invasion \cite{Thompson05,Nieto13,Yang13}. However, EMT is not the only process that may favor metastatic dissemination. 

One complementary process that may help drive the plasticity of collective cell migration is cellular jamming and unjamming \cite{Angelini11,Sadati13,Garcia15}. Recent experiments suggest that the motion of cells in tissues may be understood in terms of physical laws and parameters typically employed to study the transition between amorphous solid and liquid states of inert materials. Within this framework, epithelial cell monolayers below confluence exhibit liquid-like dynamics. As the cell density is increased due to proliferation, cellular displacements are progressively inhibited, and cells become increasingly caged by their neighbors in a glassy or jammed state \cite{Angelini11,Puliafito12} that shares many similarities with molecular or colloidal glasses \cite{Berthier11}.

Notably, the transition to a jammed, arrested state has been proposed to ensure the proper development of elasticity in mature epithelial tissues. Conversely, monolayer unjamming is needed whenever a tissue must adapt to changes or perturbations of its physiological, homeostatic state. 
Decreasing density is not the only way to cause unjamming. Recent experiments showed that an increase of cell-cell adhesion due to mechanical compression \cite{Park15} or to perturbation of endocytic processes \cite{Malinverno17} also leads to unjamming. This gateway to collective motility can be termed Jamming-to-Unjamming transition (JUT) \cite{Park16} and it may be exploited by tumors for interstitial dissemination \cite{Malinverno17}. In contrast with EMT that requires a partial of full rewiring of genetic programs and cell identity, small changes in biomechanical parameters are predicted to promote the JUT and associated collective migration. 

Theoretical models and numerical simulations play a fundamental role in understanding and guiding experiments. Particle-based models in which motile cells are treated as a collection of self-propelled entities are widely used to describe the dynamics of dense cell collectives \cite{Marchetti16,Mehes14,Smeets16,Basan13} and predict jamming as a function of cell density \cite{Fily14, Berthier14}.
Such particle-based models describe well the change of the mechanical properties of monolayers approaching confluency from a subconfluent state \cite{Henkes11}, but their effectiveness in capturing the dynamics of dense, confluent monolayers is limited. In confluent monolayers cell shape  encodes quantitative information about intercellular  interactions. This notion was demonstrated in a series of papers, in which a monolayer was modeled as a tessellation of the plane made of polygonal tiles \cite{Bi14,Bi15a,Bi15,Loza16,Barton2016}. The tissue packing fraction is unity and cells are parametrized in terms of their area and perimeters. The mechanical properties of the tissue are described by the  well established shape energy proposed in the context of the vertex model that has been used successfully to model the development of the fruit fly embryo \cite{Farhadifar07}. This model captures the JUT observed in experiments \cite{Park15} in terms of a geometric parameter that encodes the interplay between cortical tension and cell-cell adhesion \cite{Bi15a}. The addition of cell motility through a self-propelled Voronoi (SPV) model demonstrated that the JUT can be additionally tuned by cell speed and the persistence of single-cell dynamics~\cite{Bi15}. None of the current models, however, is capable of accounting for a striking set of experimental observations \cite{Malinverno17}, in which the elevation of RAB5A, a master regulator of endocytosis,  induces large-scale directed migratory patterns, which resembles the onset of flocking in other living systems \cite{Vicsek12}.

\begin{figure}[!t]
\centering
\includegraphics[width=.5\textwidth]{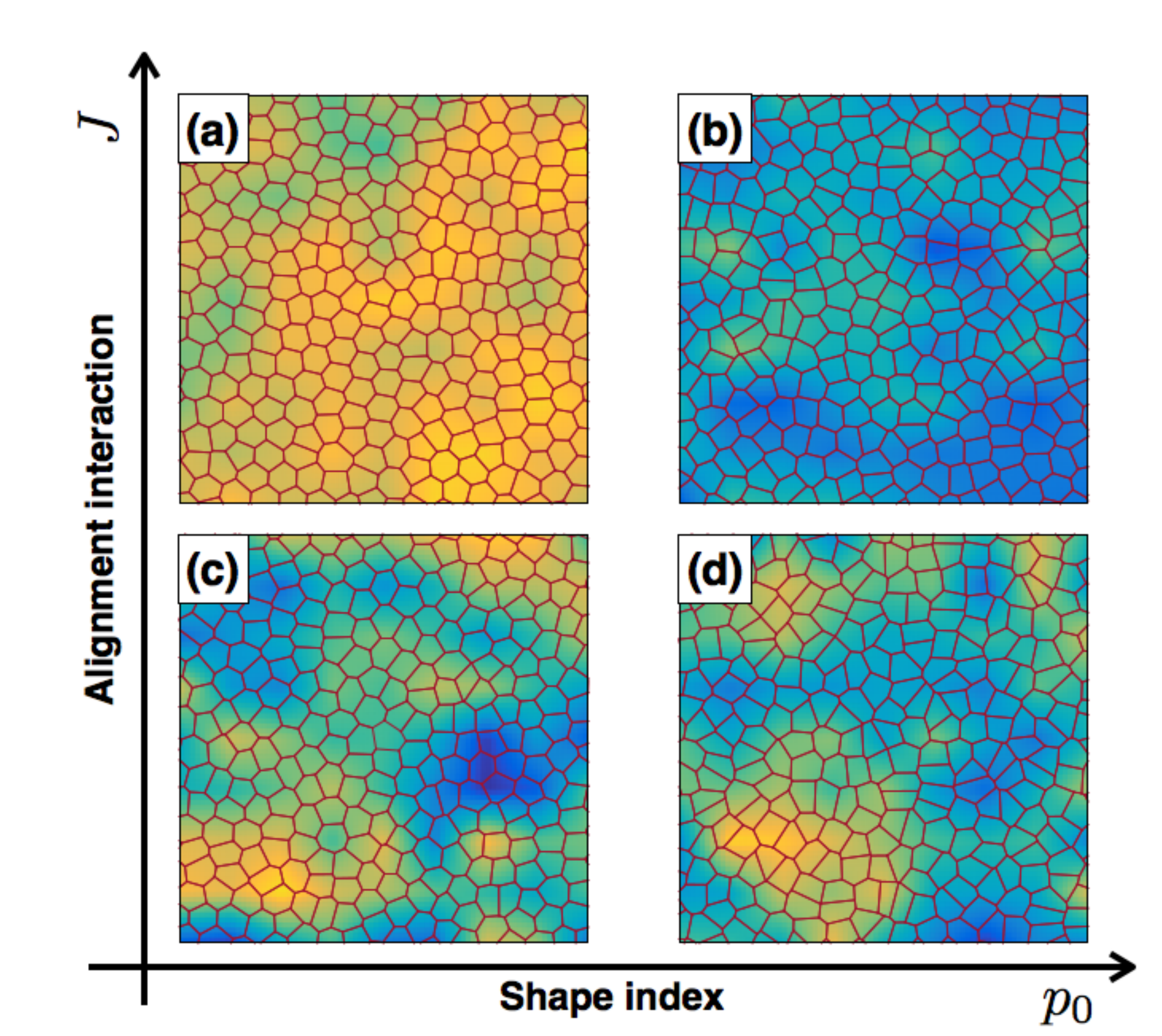}
\caption{
Four distinct dynamical phases. If the alignment interaction is strong, flocking states are observed, both solid (a) and liquid-like (b). For weak polar coupling between cells the system is either in a stationary  solid (c) or stationary liquid (d) phase. The heat map represents the cosine of the angle of the instantaneous velocity field with respect the 
horizontal axis: when the color is uniform all cells migrate coherently along the same direction. 
}
\label{fig:snap}      
\end{figure}

In addition, there is significant experimental evidence that cells alter their polarization and directional motion due to interactions with surrounding cells~\cite{Carmona-Fontaine08,Petrie09,Friedl17}.
Although there are many ways to model these types of interactions, we extend the SPV model~\cite{Bi15} by adding a simple and tractable interaction that tends to locally align cell polarization. Though simple, this modification yields a rich phase diagram, where in addition to the non-migratory liquid and solid states, flocking solid and flocking liquid phases also emerge (see Fig.~\ref{fig:snap}). 
Remarkably,  alignment not only induces coherent motion at large scales, but it also affects the structural properties of the tissue by promoting solidification and enhancing the scale of collective cellular rearrangements that occur when the solid is approached from the liquid side. The results of our model suggest that the reawakening of motility observed experimentally \cite{Malinverno17} can be understood as the result of a simultaneous increase of cell-cell adhesion and of reorientation efficiency of the cell polarization direction along the local migration velocity. 

%

\section*{Model and Methods}
\begin{figure}[!t]
\centering
\includegraphics[width=.5\textwidth]{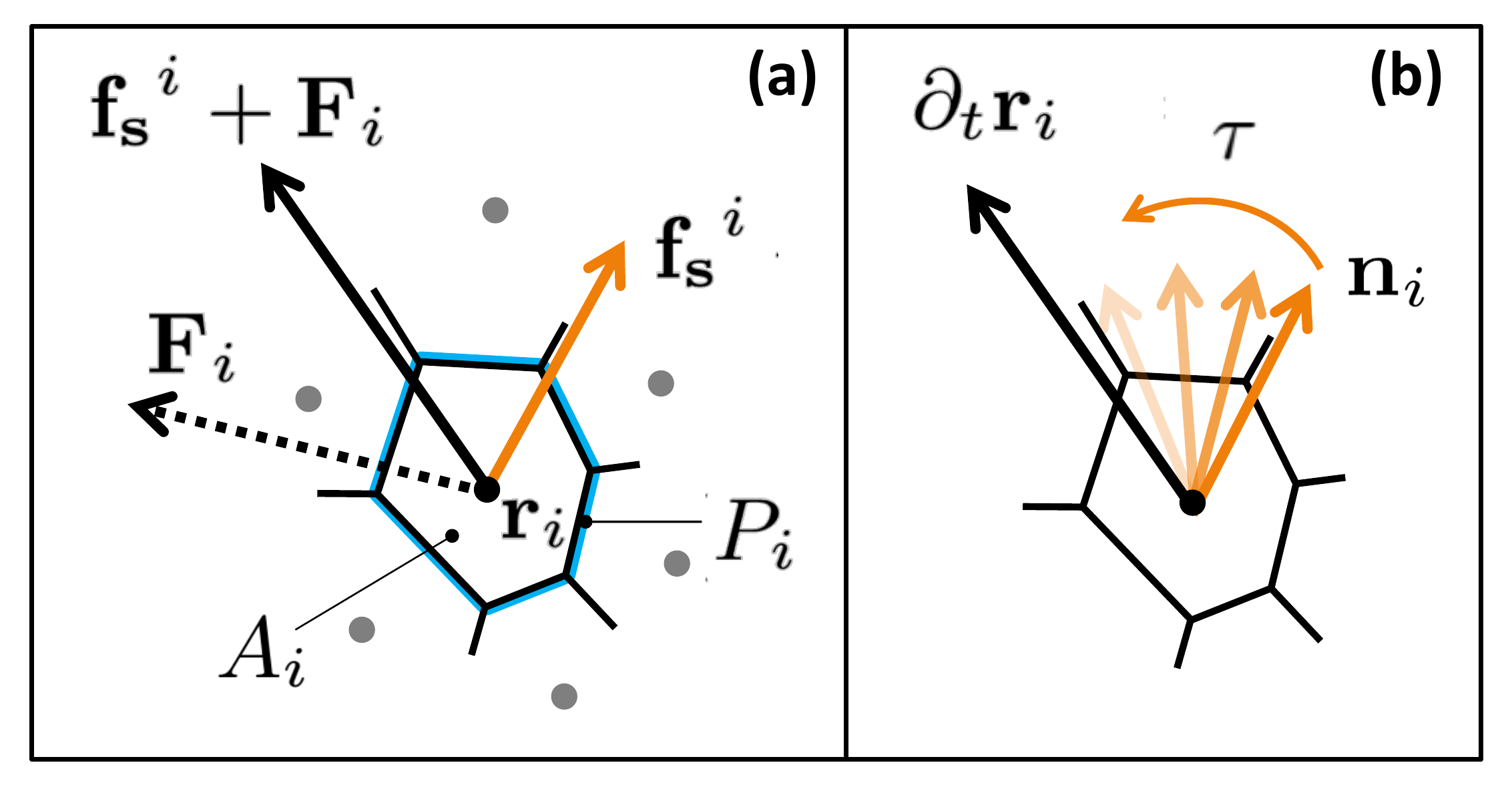}
\caption{
Schematic representation of the model. (a) Each cell is a poligon obtained by the Voronoi tessellation of initially random cell positions $\rv_i$, characterized by the area $A_i$ and the perimeter $P_i$ of the polygon.
The cell experiences a force $\Fv_i=-\nabla E$ due to its neighbors and an internal propulsive force  $\fs^i$ along the direction $\mathbf{n}_i$ of its polarization (Eq. \ref{EQ_motion}).
(b) An active orientation mechanism aligns each cell's polarization with its migration velocity with a characteristic response time $\tau=J^{-1}$ (Eq. \ref{angle}).}
\label{fig:sk}      
\end{figure}

The SPV model describes a confluent monolayer as a network of polygons covering the plane~\cite{Bi14,Bi15a,Bi15}. Each cell is characterized by its position $\rv_i$ and cell shape as determined by the Voronoi tesselation of all cell positions (Fig.~\ref{fig:sk}). As in the vertex model~\cite{Honda78}, cell-cell interactions are determined by an effective tissue energy~\cite{Honda78,Staple10,Farhadifar07,Nagai01,Hufnagel07,Wang12}. 
\BEA\label{energy}
E &\equiv& \sum_i \left[ K_A(A_i-A_0)^2 + K_P(P_i-P_0)^2\right]\;,
\EEA
with $A_i$ and $P_i$ the cross-sectional area and the perimeter  of the $i$-th cell, and   $K_A$ and $K_P$ area and perimeter stiffnesses.
The first term,  quadratic in the fluctuations of the cell area around the target value $A_0$, arises from the constraint of incompressibility in three dimensions and encodes bulk elasticity. 
The second term, quadratic in the deviation of cell perimeter from the target value $P_0$, represents the competition between  active contractility in the actomyosin  cortex and cell-cell adhesion, resulting in an effective boundary tension proportional to $P_0$. 
 We consider $N$ cells in a square box of area $L^2$ with periodic boundary conditions. In the following, we set both the average cell area $\bar{A}=L^2/N$ and the target area $A_0$ equal to one, $\bar{A}=A_0=1$, though changing $\bar{A}$ has no effect on the cell dynamics~\cite{Yang2017}. The system is initialized with random initial positions for the $N$ cells.
The configurational energy in Eq. \ref{energy} has been extensively used in the past to model biological tissues, but only recently it has been shown that this simple model exhibits a rigidity transition that takes place at constant density and it is controlled by a single non-dimensional parameter, the  target shape index $p_0 \equiv P_0/\sqrt{A_0}$~\cite{Bi15a}.
In our model, we assume that cell proliferation is negligible on the time scales of interest, as experimentally shown in \cite{Malinverno17}.
%
%
%

Each cell is additionally endowed with motility described by a self-propulsive force $\fs^i=f^0 \mathbf{n}_i$ of fixed magnitude $f_0=v_0/\mu$, with $v_0$ the cell motility and $\mu$ a mobility, pointing along the direction $\mathbf{n}_i=\left( \cos \theta_i, \sin \theta_i \right)$ of cell polarization. Assuming overdamped dynamics, the equation of motion of cell $i$ is
\BEA\label{EQ_motion}
\p_t \rv_i &=& \mu \left( \fs^i + \Fv_i \right) \;, 
\EEA
with 
 $\Fv_i = -\nabla_{\rv_i} E$ the force arising from the tissue energy. 
In Ref.~\cite{Bi15} the direction of polarization was assumed to be determined entirely by rotational noise, independent of the state of neighboring cells.  Many cell types, however, are known to sense mechanical and biochemical stimuli from neighboring cells and actively respond by adjusting their polarization~\cite{Stramer16}. 
Following earlier work~ \cite{Sza06,Henkes11},  we model these interactions as an active feedback mechanism at the single cell level that tends to align each cell's polarization with its migration velocity, which is in turn controlled by interactions with other cells. The polarization dynamics is then governed by the equation
%
%
%
\BEA\label{angle}
\p_t \theta_i&=& - J \sin(\theta_i - \phi_i) + \eta_i \; , 
\EEA
where $\phi_i$ is the direction of the cell velocity, $\p_t \rv_i=\vv_i= v_i \left( \cos \phi_i , \sin \phi_i \right)$, and $\eta_i$  a white noise, i. e., $\langle \eta_i(t) \rangle = 0$ and  $\langle \eta_i(t) \eta_j(s) \rangle = 2 D_r \delta_{ij} \delta(t-s)$. The angular dynamics is controlled by the interplay of rotational diffusion at rate $D_r$ and alignment at rate $J$, whose inverse $\tau=J^{-1}$ is the response time required by the cell to reorient its polarization in the direction along which it is pushed by its neighbors. 
%
%
In the following, we use  $\sqrt{A_0}$ as the unit length and $(\mu K_A A_0)^{-1}$ as the unit time \cite{Bi15}. Additionally, we set $K_P/(K_A A_0)=1$, $f_0/(K_A A_0^\frac{3}{2})=1$, $D_r/(\mu K_A A_0)=0.5$. 
The free parameters are thus the (dimensionless) alignment rate $J$ and the target shape index $p_0$. For $J=0$ and $v_0=0$ our model is related to the vertex model in Ref.~\cite{Bi15a}, while for $J=0$ and finite $v_0$ we obtain the SPV model of Ref.~\cite{Bi15}. 


To study the solid-liquid transition, we use the mean-square-displacement $MSD(t)=N^{-1}\langle \sum_i [\rr_i^\prime (t) - \rr_i^\prime (0)]^2 \rangle $ evaluated in the reference frame of the center of mass $\rr_{CM}=N^{-1} \sum_i \rr_i$, with  $\rr_i^\prime=\rr_i - \rr_{CM}$. The normalized self-diffusivity $D_{self} \equiv \lim_{t\to\infty} \frac{MSD(t)}{4 t D_0}$ is a dynamical order parameter for the onset of rigidity, which can also be identified via a structural order parameter given by the cellular shape index \cite{Bi15}, $q= \langle P_i/\sqrt{A_i}\rangle$, where the brackets denote an average over cells. The transition line $D_{self}\leq 10^{-3}$ corresponds to $q=3.813$. When $v_0=0$, the rigidity transition occurs for $p_0=p_0^*=3.81$ \cite{Bi15a}. 

We quantify the emergence of flocking by using the Vicsek order parameter $\varphi \equiv N^{-1} \left\langle | \left(\sum_i \vv_i / |\vv_i|\right)| \right\rangle$, where the angular brackets indicate the average over trajectories. This quantity vanishes when cells are moving in random directions and attains a value of $1$ when all cells coordinate their motion. The susceptibility $\chi_{\varphi} = \langle \left( \varphi(t) - \langle \varphi \rangle \right)^2 \rangle $ exhibits a maximum at the flocking transition, which we use to separate flocking from non-flocking states.

\section*{Results}

\begin{figure}[!t]
\centering
\includegraphics[width=.5\textwidth]{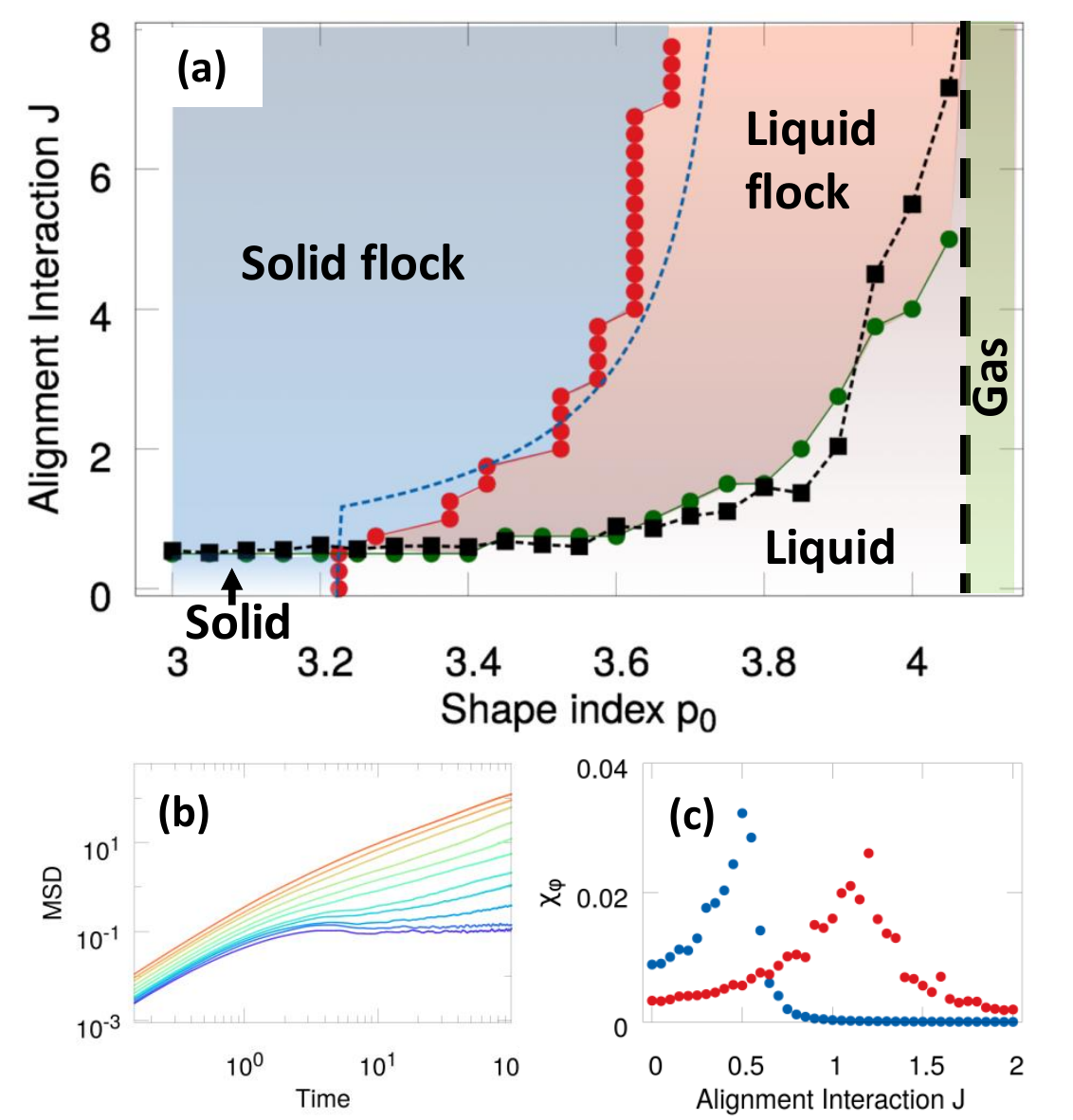}
\caption{Phase diagram. (a) Different phases in the $(p_0,J)$ plane. The solid/liquid transition line (red circles) is obtained from the vanishing of $D_{eff}$ and the flocking transition line (green circles) corresponds to the peak in the susceptibility $\chi_{\varphi}$. The dashed blue curve is the theoretical prediction $J_c(v_0,p_0)$ given in ~\eqref{Jc}. The black squares (the dashed line is a guide to the eye) are the estimate for $J_{flock}(p_0)$ in terms of the numerically calculated cage lifetime $\tau_{cage}$ at $J=0$. 
The vertical dashed black line marks the transition to a gas-like state, observed for $p_0\gtrsim4.2$, where cells behave almost independently and flocking can not occur. 
(b) The mean square displacement for $J=2.0$ for a range of $p_0\in[3.4,4]$ across the liquid/solid transition (curves from red to violet). (c) The susceptibility $\chi_{\varphi}$ for $p_0=3.1$ (blue circles, solid) and $p_0=3.7$ (red circles, liquid).} 
\label{fig:pd}      
\end{figure}

\begin{figure}[!h]
\centering
\includegraphics[width=.45\textwidth]{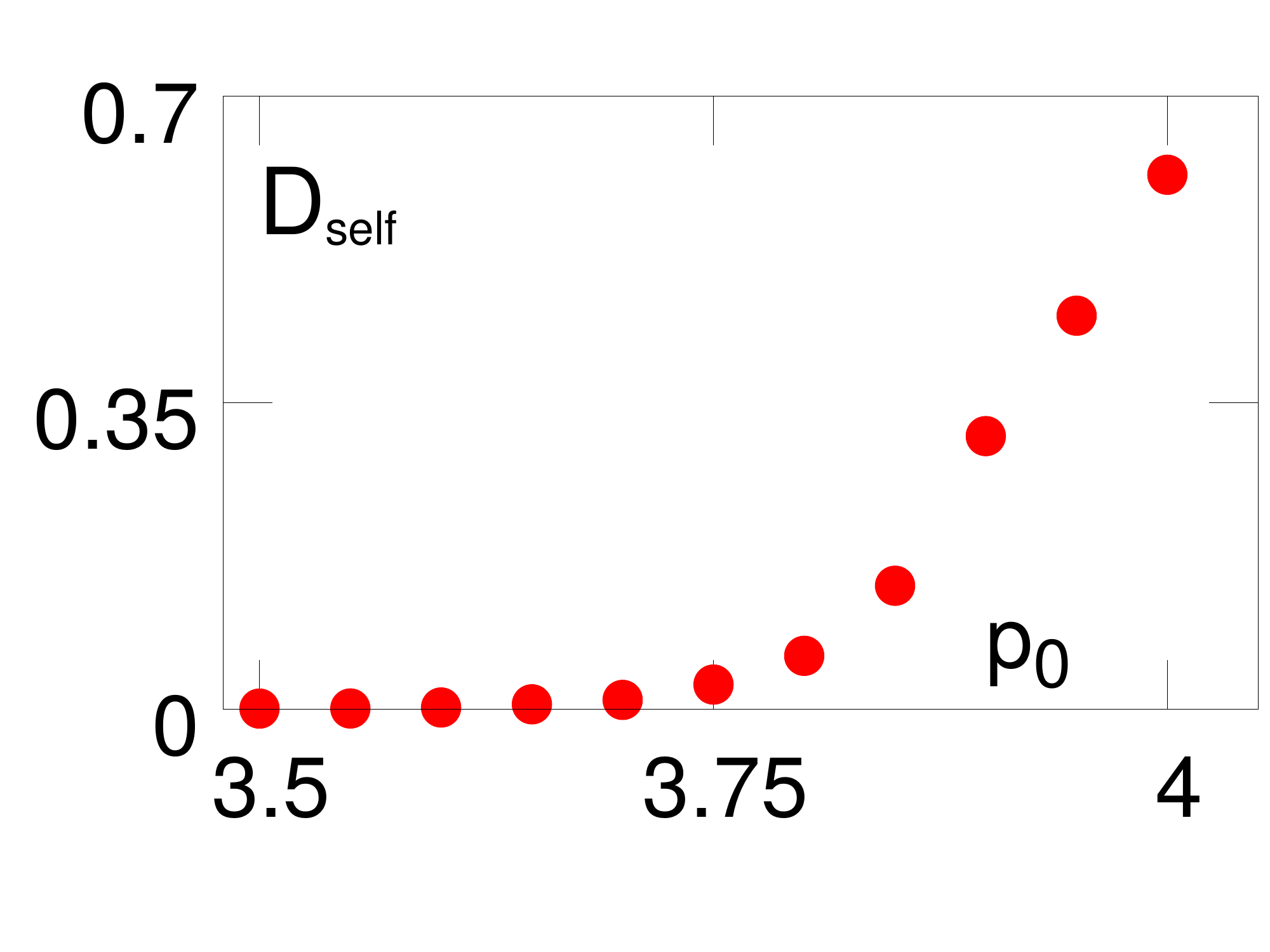}
\caption{Rigidity transition. Normalized self-diffusivity $D_{self} \equiv \lim_{t\to\infty} \frac{MSD(t)}{4 t D_0}$ as a function of the shape index $p_0$ for $J=2.0$. $D_{self}$ is a dynamical order parameter for the onset of rigidity. When $D_{self}<10^{-3}$ the system is considered to be in solid state.}
\label{fig:deff}
\end{figure}

As shown in Fig.~\ref{fig:snap}, we find four distinct phases by varying the alignment rate $J$ and the target shape index $p_0$: (a) a stationary solid with vanishing $D_{self}$, corresponding to the absence of cellular rearrangements, and $\varphi=0$; (b) a stationary liquid with finite $D_{self}$ and vanishing mean motion ($\varphi=0$);  (c) a flocking liquid where cells flow collectively ($D_{eff}$ and $\varphi$ are both finite); and (d) a flocking solid where the tissue migrates as a unit ($\varphi$ finite), while maintaining its rigidity. 
 A phase diagram for the system is shown in Fig.~\ref{fig:pd}-a. The solid/liquid transition (red circles) has been determined by examining the MSD that evolves from diffusive to saturated with increasing $p_0$ (Fig.~\ref{fig:pd}(b)), resulting in the vanishing of the long time diffusivity $D_{self}$ (see Fig. \ref{fig:deff}). The line separating the non-flocking from the flocking phases (green circles) corresponds to the peak in the susceptibility shown in panel (c). The dashed blue line and the black squares are theoretical estimates described below.
It is clear from the phase diagram that there is a subtle interplay between the structural and mechanical properties of the tissue as controlled by the shape parameter $p_0$ and the onset of collective migration driven by the alignment $J$. 
In the following, we discuss how alignment impacts the mechanical properties of tissue by promoting solidification and enhancing collective cellular rearrangements at the liquid-solid transition.

%

%

\section*{Flocking promotes solidification}
It is evident from the behavior of the red line separating the flocking liquid from the flocking solid in Fig.~\ref{fig:pd}  that, counterintuitively, alignment promotes solidification. We show in this section that this effect can be understood by a simple argument that also provides an estimate for the liquid-solid transition line in the flocking region,
$p_c(v_0,J)$.  Briefly, as suggested in Ref.~\cite{Bi15}, fluidification can be understood qualitatively in terms of an ``effective temperature'' that allows cells to rearrange by overcoming the energy barriers associated with $T_1$ transitions. When cell alignment  is faster than rotational diffusion (i.e., $J\gg D_r$)  cells can move coherently without being disrupted by noise, which results in a lower effective temperature, and therefore promotes solidification. 
%

To flesh out this argument we first recall that in Ref. \cite{Bi15a} it was shown that in a static vertex model described by the tissue energy of Eq. \ref{energy}, with $v_0=0$, the transition from solid to liquid is associated with the vanishing of the mean energy barriers $\Delta E$  for $T_1$ transitions and that these barriers scale as $\Delta E \propto p_0^*-p_0$  when the target shape index $p_0$ approaches its critical value $p_0^*=3.813$ from the solid side. 
Following \cite{Bi15}, we assume that in the SPV model the effect of cell motility can be accounted for through an effective temperature $T_{eff}$ controlled by the fluctuations in cells  positions  that allow each cell to locally explore its energy landscape. In the absence of cell-cell alignment ($J=0$) this argument was used in \cite{Bi15} to obtain an excellent fit to the liquid-solid transition line at finite $v_0$ using $T_{eff}=c v_0^2$, with $c$ a dimensionful fitting parameter. Here we make the argument more precise and generalize it to finite $J$.

In the gas phase, where both interactions and alignment can be neglected, an exact calculation of the mean-square displacement of a single cell yields the identification $k_BT_{eff}^g=v_0^2/\mu D_r$~\cite{Fily12}. In the solid, the cells are caged by their neighbors. Considering first  $J=0$, caging can be modeled by assuming that each cell is tethered to a spring of force constant $k$. An exact calculation of the mean square displacement of a tethered motile cell in the presence of orientational noise (see SI) yields $\lim_{t\rightarrow\infty}MSD(t)=v_0^2/[\mu k(\mu k+D_r)]$. Comparison with the corresponding result for a Brownian particle tethered to a spring, $\lim_{t\rightarrow\infty}\langle[\Delta\mathbf{r}(t)]^2\rangle_{th}=k_BT/k$ suggests the identification of  an effective temperature $k_BT_{eff}^s=v_0^2/[\mu(\mu k+D_r)]$ \cite{Maggi14}.
By assuming that the transition is controlled by the balance of the energy barrier and this effective thermal energy, $\Delta E\sim p_0^*-p_c(v_0,J=0)\sim T_{eff}^s$, we obtain a critical line for the solid-liquid transition 
%
$p_c(v_0,J=0)=p_0^*-v_0^2/[\mu(\mu k+D_r)]$,
%
 consistent with the result of  \cite{Bi15}.  As discussed in \cite{Bi15} this argument works best at large $D_r$, where the effect of rotational noise resembles that of thermal fluctuations. 

A similar argument accounts for the role of alignment. For $J \ll D_r=0.5$, the alignment interaction is ineffective and does affect the location of the solid-liquid transition. For large $J$, however, the system is in a solid flocking state, characterized by a finite mean velocity $\vec{v}=\bar{v}(\cos\bar\phi,\sin\bar\phi)$. We consider again a single cell tethered to a spring of force constant $k$ to describe caging by neighbors, but also moving at mean velocity $\vec{v}$. Fluctuations about this ordered state are mainly transverse to the direction of mean motion. Treating such fluctuations as small, the mean square displacement of such a solid flocking cell is given by (see SI for details) $\lim_{t\rightarrow\infty}MSD(t)=v_0^2D_r/[\mu kJ(\mu k+J(1-v_0/\bar{v})]$.  Assuming $\bar{v}\sim v_0$, the corresponding effective temperature is $T_{eff}^f=v_0^2 D_r/(\mu^2 kJ)$. 
Equating again this
 thermal energy to  the energy barriers for $T1$ transition,
we obtain an estimate for the transition line between solid and liquid flocks as 
$p_c(v_0,J\gg D_r)\sim p_0^*-v_0^2D_r/(\mu^2 k J)$.
This yields the transition from flocking solid to flocking liquid as
\begin{equation}
J_c(v_0,p_0)\sim \frac{v_0^2D_r}{\mu^2 k}~\frac{1}{p_0^*-p_0}\;,
\label{Jc}
\end{equation} 
which provides a good fit to the data with $k=0.85\pm 0.03$ (dashed blue line in Fig.~\ref{fig:pd}).
%
%
%
%

\section*{Flocking requires slow structural rearrangements}
\begin{figure}[!t]
\centering
\includegraphics[width=.5\textwidth]{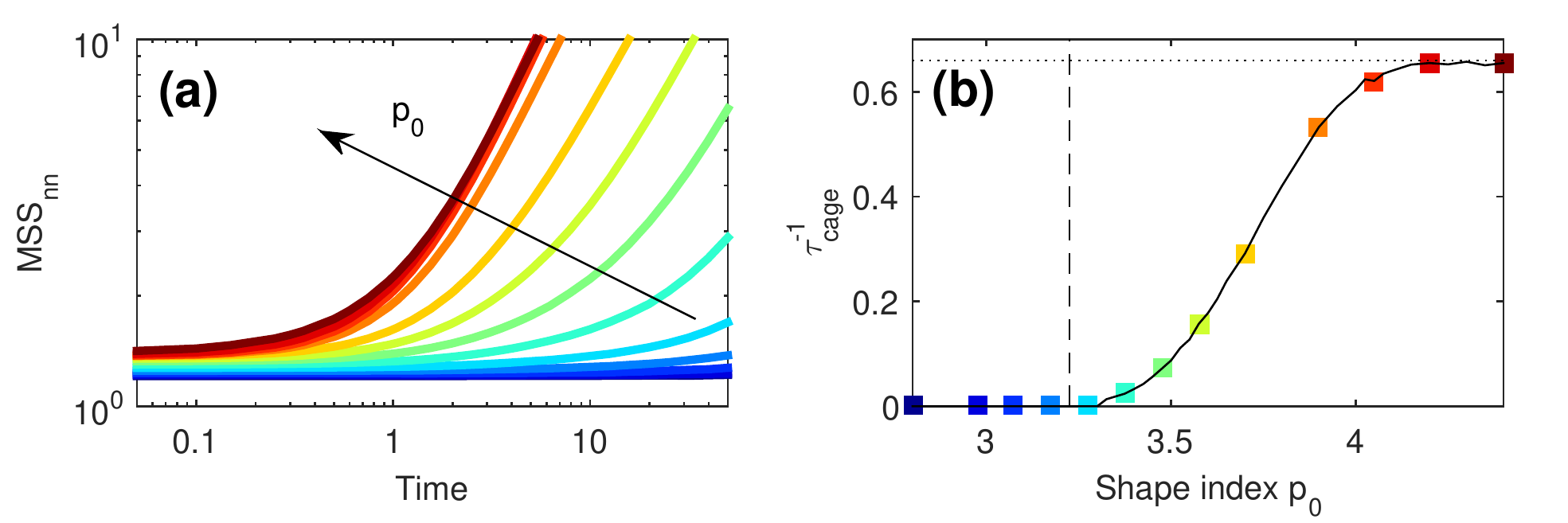}
\caption{Caging for  $J=0$. (a) The neighbors mean-square separation $MSS_{nn}(t)$ as a function of the the time increment $t$ becomes constant as $p_0$ is decreased across the liquid-solid transition, showing the onset of caging. (b) The inverse cage lifetime $\tau_{cage}^{-1}$ at $J=0$ as a function of $p_0$ calculated as described in the text. The vertical line denotes the critical value $p_0^*$ of the $J=0$ rigidity transition, while the horizontal dotted line is the asymptotic value $\tau_{free}^{-1}$ attained by $\tau_{cage}^{-1}$ in the gas phase.}
\label{fig:cage_time}      
\end{figure}
The line separating the stationary from the flocking liquid (Fig.~\ref{fig:pd}, green) can be estimated by equating the time scale $\tau=J^{-1}$, with which a cell aligns its polarization along the migration direction, with $\tau_{cage}$, the lifetime of the local cages. Deep in the solid, $\tau_{cage}$ is infinite and structural rearrangements are controlled solely by the time scale $\tau_r$ for rotational diffusion. Cells will align with their neighbors provided that $\tau<\tau_r$, giving a critical value $J_{flock}(p_0\ll p_0^*)=D_r$ for the onset of flocking in the solid, independent of $p_0$ and in agreement with Fig.~\ref{fig:pd}. As the solid-liquid transition is approached, $\tau_{cage}$ becomes finite. When $J=0$, $\tau_{cage}$ can be estimated as the time over which the mean-square separation of two cells $i$ and $j$ that are in contact at $t=0$, defined as $MSS_{nn}(t)=\langle [\rr_i(t) - \rr_j(t)]^2 \rangle $, remains constant (see SI for details). The neighbors mean-square separation is shown in Fig.~\ref{fig:cage_time}(a) for $J=0$ and several values of $p_0$ spanning the liquid-solid transition. In the solid,  $MSS(t)$ is constant at all times. Upon melting, \textit{i.e.}, for $p_0 > p_0^c(v_0,J)$, $MSS_{nn}(t)$ shows an initial plateau and then starts to grow. 
The resulting inverse lifetime of the cage is shown in Fig.~\ref{fig:cage_time}(b)   as a function of $p_0$. The lifetime $\tau_{cage}$ 
reaches a constant value for $p_0\gtrsim4.2$, where $q\simeq 4.19$.  This value is close to the value $q_{rand}=4.186$ corresponding to the Voronoi tessellation of randomly distributed points, indicating that for $p_0\gtrsim4.2$ the system is essentially a gas. In this gas regime $MSS_{nn}$ still shows an initial plateau at short time that corresponds to the time $\tau_{free}=a/v_0$ taken by a cell of motility $v_0$ to travel freely a distance of the order of its size $a\sim\sqrt{A_0}$. We then define the true cage lifetime $\tau^*_{cage}$ by correcting the lifetime calculated from the $MSS_{nn}$ as $\tau^*_{cage}=\tau_{cage}-\tau_{free}$.

In the liquid, structural rearrangements can occur via both the relaxation of the local cage on time scale $\tau_{cage}^*$ and noisy reorientation on time scales $\tau_r$.  Neighbor exchanges are  controlled by the faster of the two processes. Flocking will only occur if the alignment rate $J$ is faster than the total rate $1/\tau_{cage}^*+1/\tau_r$ for neighbor exchanges, giving an estimate for the flocking transition in the liquid as $J_{flock}(p_0)=1/\tau_{cage}^*+D_r$. This prediction yields the black squares in Fig.~\ref{fig:pd} in good agreement with the phase boundary shown in green.  Finally, we note that the existence of a gas phase also explains the observed vertical asymptote in $J_{flock}(p_0)$: if cells are not interacting they cannot align their polarization vectors, no matter how rapidly the mutual alignment occurs.

\begin{figure}[!t]
\centering
\includegraphics[width=.5\textwidth]{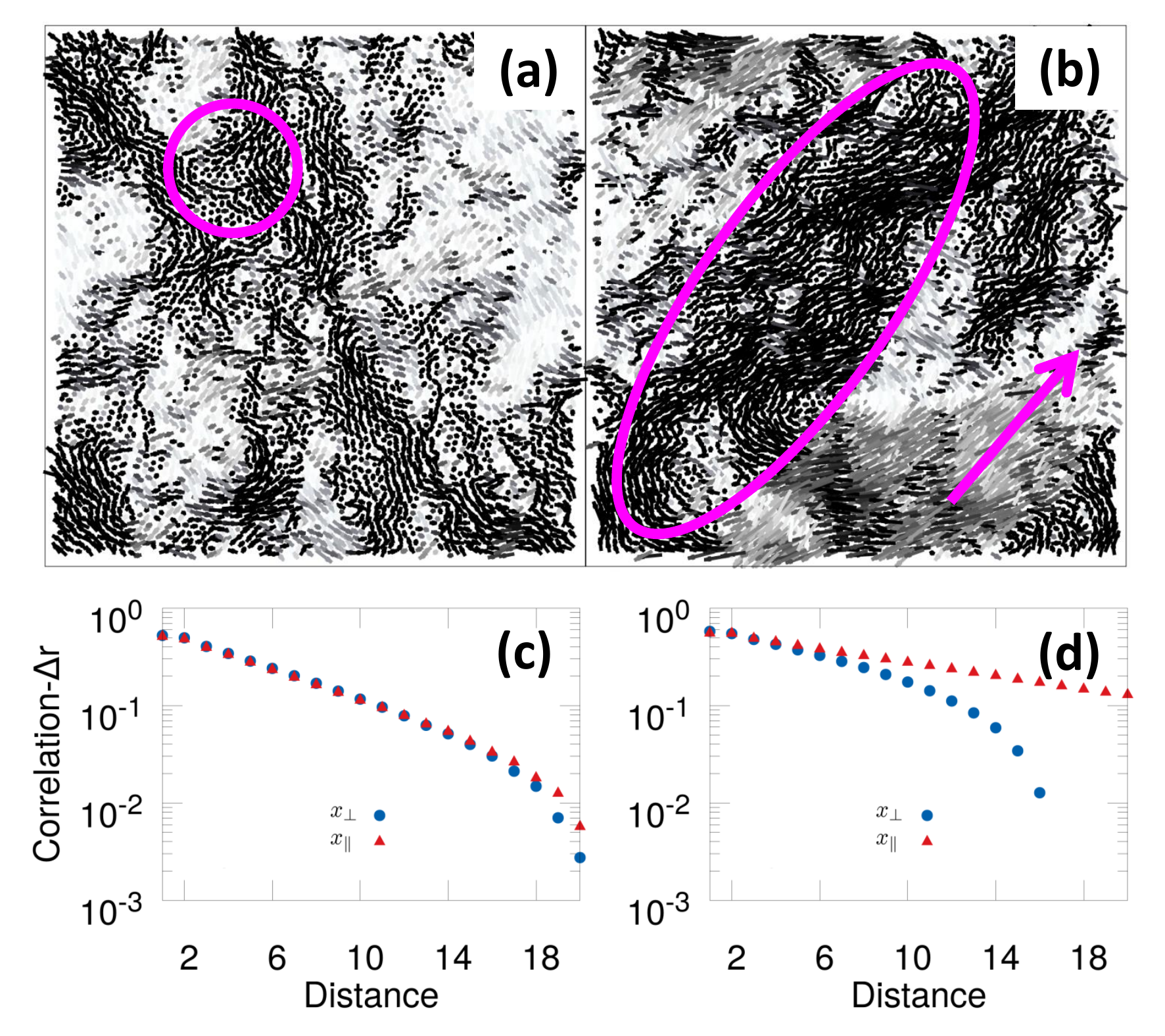}
\caption{Dynamical Heterogeneities. Maps of the displacements $\Delta \rr_i$  averaged over a time $\tau_{\alpha} = 10^2$ for (a) $J=0$ and 
(b) $J=2$ in a system of $4900$ cells. Magenta circles are a guide to the eye highlighting the anisotropy of the collective rearrangements in the flocking state. Magenta arrow indicates the average migration direction. 
Frames (c) and (d) show the spatial correlations  
 $C(x_\parallel,0)$ (red triangles) and  $C(0,x_\perp)$ (blue circles) along orthogonal axes longitudinal ($x_\parallel$) and perpendicular ($x_\perp$) to the direction of mean motion of a given sample for $J=0$ (c) and $J=2$ (d), averaged over $10^2$ samples (see SI for details).
}
\label{fig:fig_map}      
\end{figure}

\section*{Flocking enhances dynamical heterogeneities at the jamming transition}
As in many glassy systems, the liquid-solid transition in the SPV model reported in \cite{Bi15} is accompanied by slowing down of the structural relaxation when the solid is approached from the liquid side, with collective rearrangements that take the form of swirling motion on growing scales, known as dynamical heterogeneities. In particulate glasses, these heterogeneities can be difficult to see and characterize, and so a four-point correlation function $\chi_4$ is typically used to identify length and timescales. Here, the presence of alignment interactions strongly slows the dynamics and enhances the anisotropy of collective rearrangements so they resemble local flocks. This is evident in the maps of cellular displacements shown in Fig.~\ref{fig:fig_map}.
These show cellular displacements $\Delta \rr_i = \left\langle \left[ \rr_i^\prime(\tau_\alpha + t) - \rr_i^\prime(t) \right] \right\rangle_t $ time-averaged over of the order of the structural relaxation time $\tau_\alpha$ calculated in the SI. 
For $J=0$ (panel a), the rearrangements (highlighted by the magenta dashes) are small isotropic swirls, while for $J=2$ they become anisotropic flocks (panel b).
It is worth noting that 
the displacements are computed in the center of mass frame, hence
the local flocks are not due to mean motion,
but to heterogeneities in the local rearrangements. 
To quantify the dynamical heterogeneities and highlight their anisotropic structure in the flocking state, we have evaluated the spatial correlation of cell displacements $C(x_\parallel,x_\perp)$ along directions longitudinal ($x_\parallel$) and transverse ($x_\perp$) to that of mean motion.
In absence of alignment ($J=0$, Fig. \ref{fig:fig_map}(c)), the correlation is isotropic, but becomes strongly anisotropic in the flocking state  ($J=2$, Fig. \ref{fig:fig_map}(d)).
%

\begin{figure}[!t]
\centering
\includegraphics[width=.45\textwidth]{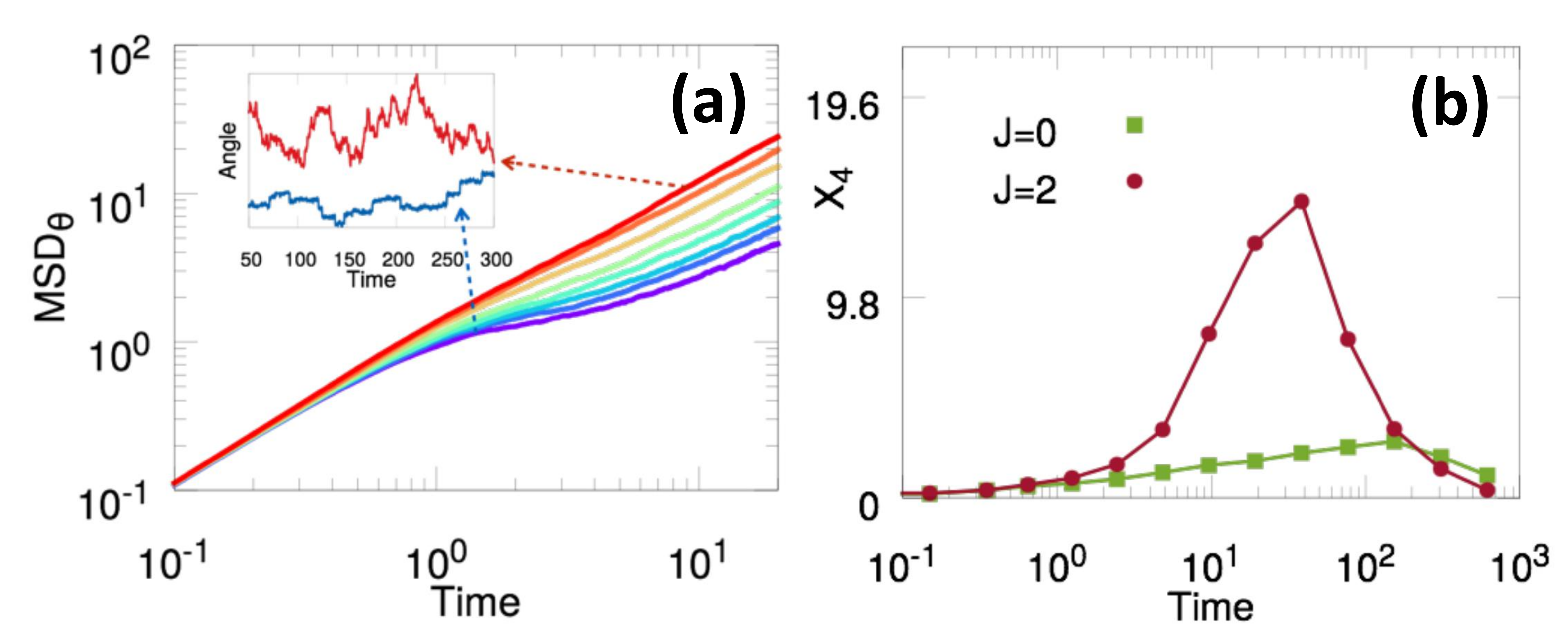}
\caption{Mean-squared angular displacement $MSD_\theta$. (a) $MSD_\theta$ of the polarization angle $\theta$ for $J=2.0$ and $p_0\in[3.0,3.5]$, i. e., 
in the flocking regime. $MSD_\theta$ becomes subdiffusive on short time scales.
Also shown are two typical angular trajectories: for large $p_0$ (red curve) $\theta(t)$ performs a random walk while for small $p_0$ (blue curve) the 
dynamics of $\theta(t)$ is characterized by jumps local and vibrations.
(b) $\chi_4(t)$ approaching the liquid-solid transition ($p_0-p_0^c/p_0^c\sim 10^{-3}$) for 
$J=0$ (green squares) and $J=2$ (red circles) in a system of $N=4900$ cells.}
\label{fig:theta}      
\end{figure}

The timescale associated with correlated motion is easily identified by examining the angular displacements of cell polarization, quantified  by
the angular mean square displacement of individual cells, $ 
MSD_\theta (t)\equiv N^{-1}\left\langle \sum_i \left[ \theta_i(t)-\theta_i(0) \right]^2 \right \rangle$.  For $J=0$, $\theta_i$ undergoes a random-walk and $MSD_\theta=2 D_r t$. 
The situation changes when the polar interaction is turned on.
In the flocking state, the $MSD_\theta$ becomes subdiffusive on intermediate scales, as evident in Fig.~\ref{fig:theta} for $p_0<3.5$.
The subdiffusive regime is due to the presence of jump processes in the dynamics 
of $\theta(t)$ that are apparent in the blue trajectory shown in the lower inset to Fig.~\ref{fig:theta}.  
%
To explore the link between subdiffusion and dynamical heterogeneity, we have computed the non-Gaussian parameter (the kurtosis of the distribution of the angular displacements)~\cite{KobGlotzer1997}
that develops a clear peak for $t \sim 10$ (see SI). Also, since the time scale of angular relaxation is an order of magnitude smaller than that of structural relaxation, the displacements due to the flocking excitations give the dominant contribution to the displacements in Fig. \ref{fig:fig_map}b. The contribution due to flocking is also clear when looking at the behavior of $\chi_4$ susceptibility shown in \ref{fig:theta}b.


%


%

%


\section*{Discussion}
In this work, we describe a minimal model for collective migration in biological tissues. Our model treats a confluent cell monolayer as a Voronoi tessellation of the plane and encodes mechanical properties of the cells, such as intracellular adhesion, cortical tension, and motility. Motivated by experiments at both the tissue and cellular scales, we introduce a polar interaction mechanism similar to the one leading to flocking in other active matter systems [\cite{Vicsek95,Sza06,Vicsek12}] that captures the feedback between local dynamics and cell polarization. By tuning the strength of the polar interaction and the preferred perimeter of the cells, we find a rich phase diagram with four phases. At low polar interaction strengths, we find standard liquid and amorphous solid phases. Increasing interactions, we find two new phases: an amorphous flocking solid, and a flocking liquid, both exhibiting collective directed motion.

Our findings suggest that polarization alignment yields global migration and can promote solidification. 
Remarkably, our phase diagram captures the JUT observed in recent experiments on epithelial monolayers, where overexpression of the endocytic protein RAB5A triggers the onset of directed collective motion in an otherwise quiescent monolayer and promotes local fluidization \cite{Malinverno17}. Cell migration patterns in this state are compatible with the flocking liquid state predicted by our model in the regime where both the polar interaction strength and the target perimeter $p_0$ are large. 

Can the JUT observed in experiments \cite{Malinverno17} and captured by our polar-SPV model be an alternative mechanism to EMT for enhancing cell migratory capabilities in cancer? 
It has been suggested the EMT and its inverse MET are important in cancer invasion, as EMT may allow broad dispersion of cancerous cells while MET facilitates the growth of secondary tumor foci in distant microenvironments~\cite{Nieto13}. These ideas are backed by single cell studies, where it has been shown that sessile epithelial cells can change to locomotory mesenchymal phenotypes with additional cancer stem cell traits.
On the other hand, multicellular studies do not identify a clear connection between EMT-MET and cell dispersion. For example, it is possible for purely mesenchymal cells to migrate collectively~\cite{Theveneau10}, similar to the flocking fluid state in our model. Solid tumors can also disseminate while maintaining their epithelial mechanical properties (including tight cell-cell adhesion), similar to the flocking solid state. Moreover, contact forces between cells have shown to be essential for collective migration in mesenchymal cells~\cite{Theveneau10}, which is consistent with our model prediction that flocking phases only occur when the alignment with forces due to neighbors is strong.

Due to difficulties in quantifying cell polarization in multicellular sheets, there are no direct measures of how a cell's polarization changes in response to changes in its local environment. An exciting direction for future work is the study of subcellular stuctures or intercellular markers for cell polarization in multicelluar monolayers to correlate those with cell shapes and interfacial tensions and test the hypothesis that cells polarize according to mechanical forces generated by neighboring cells. If so, it may even be possible to extract the time constant $J^{-1}$ associated with this alignment.

Finally, there is another important difference between EMT and JUT. The EMT requires metastable changes in cell identity and genetic transcriptional programs. While these changes have been well documented in experimental models, evidence that they occur in real tumors is largely missing and debated \cite{Theveneau10}. In contrast, JUT requires only small and relatively simple changes to cell properties, such as the actomyosin contractility, the level of adhesion molecules at the cell boundary, or even the timescale over which cells alter their polarization in response to neighboring forces. JUT thus appears as a flexible, reversible strategy that may facilitate both dissemination and new tumor growth without requiring rewiring of a cell's genetic makeup. 



\section*{Acknowledgments}
We thank Matthias Merkel for developing the code used in some of the simulations. We acknowledge support from the Simons Foundation Targeted Grant in the Mathematical Modeling of Living Systems 342354 (MP and MCM), Simons Foundation grants 446222 and 454947 (MLM), the Syracuse Soft Matter Program (DB, MP, MLM and MCM), the National Science Foundation DMR-1305184 (MCM) and DMR-1352184 (MLM), National Institutes of Health R01GM117598-02 (MLM), the Italian Ministry of University and Scientific Research (MIUR) Project RBFR125H0M (FG, MM and RC), from Regione Lombardia and CARIPLO foundation - Project 2016-0998 (FG, MM and RC). Computing infrastructure support was provided by NSF ACI-1541396.

\section*{Supplementary Information}

\subsection*{Model simulation details}
We performed numerical simulations  by integrating the equations of motion with an Euler method for $N_t=2^{17}-2^{22}$ steps with integration time step $\Delta t=10^{-2}$. The phase diagram was investigated performing $N_J\times N_p$ simulations, $N_J=22$ with $J\in[0,8]$ and $N_{p_0}=32$, $p_0\in[2.9,4.1]$. The robustness of our results against finite size effects was evaluated by considering different systems sizes, namely $N=100,256,400,900,4900$.

\subsection*{Mean square displacement in the solid phases}

In the solid we model each cell $i$ caged by its neighbors as
a point particle tethered to a spring of elastic constant $k$.
Considering first the isotropic solid, the dynamics of the
fluctuations $\delta \rr_i=\rr_i - \rr_i^0$ of a caged particle around its mean position, $\rr_i^0$, is goiverned by the equations
\BEA  \label{j0}
\delta \dot{\rr}_i &=&  v_0 \mathbf{n}_i - \mu k \delta \rr_i\;, \\ \nn
\dot{\theta}_i &=& \eta_i\;.
\EEA  
The mean square displacement can be calculated analytically [57], with the result
\BEQ 
MSD(t)= \frac{v_0^2}{2 \mu k} \frac{1 - e^{-\mu k t} - \frac{\mu k}{D_r}\left( 1 - e^{-D_r t}\right)  }{1 - (\frac{\mu k}{D_r} )^2} \;,
\EEQ 
and the long-time  limit
\BEQ 
\lim_{t \to \infty} MSD(t) = \frac{v_0^2}{\mu k ( \mu k + D_r)} \; .
\EEQ 

We now want to examine the mean-square displacement deep into the solid flocking state, in the limit large $J\gg D_r$.
We model again an individual cell as a particle tethered to a spring of force constant $k$ due to caging from the neighbors, but also moving at the mean velocity  $\bar{\vv}=\bar{v}(\cos\bar{\phi},\sin\bar{\phi})$ of the flock.
We orient  the $x$ axis along the direction of mean migration, corresponding to $\bar{\phi}=0$, and let $\vv_i=\bar{\vv} + \delta \vv_i$,
with $\delta \vv_i=(\dot{x}_i,\dot{y}_i)$, and $\dot{x}_i\simeq\delta v$ and 
$\dot{y}_i\simeq\bar{v}\phi_i$. Letting
 $\ee_i\simeq (1,\theta_i)$,
the equations of motion for the fluctuations are
\BEA 
\dot{x}_i &=& v_0 - \mu k x_i\;, \label{eq:x}\\ 
\dot{y}_i &=& v_0 \theta_i - \mu k y_i\;, \label{eq:y}\\ 
\dot{\theta}_i &=& -J \left(\theta_i - \frac{\dot{y}_i}{\bar{v}}\right) + \eta_i \, .\label{eq:theta}
\EEA 
Fluctuations transverse and logitudinal to the direction of mean motion are decoupled. Eliminating the angular dynamics in favor of $y_i$, Eqs.~\eqref{eq:y} and ~\eqref{eq:theta} can be recast in the form of a second order differential equation for $y_i$. At long times this reduces to
\BEQ 
\dot{y}_i = -\frac{J \mu k y_i}{\mu k + J \left( 1 - \frac{v}{v_0} \right) } + \frac{v_0 \eta_i}{\mu k + J \left( 1 - \frac{v}{v_0} \right) }\;.
\EEQ 
The transverse part of the mean-square displacement can then be immediately obtained as
\BEQ 
\langle[y_i(t)]^2\rangle=\frac{v_0^2 D_r}{J \mu k \left[ \mu k + J \left( 1 - \frac{v}{v_0} \right) \right]}\left( 1 - e^{-\frac{J \mu k }{\mu k + J (1 - \frac{v}{v_0})} t}\right) \, ,
\EEQ 
with long-time limit 
\BEQ 
\lim_{t\to\infty}\langle[y_i(t)]^2\rangle =\frac{v_0^2 D_r}{J \mu k \left[ \mu k + J \left( 1 - \frac{v}{v_0} \right) \right]} \, .
\EEQ 
Deep in the flocking state we can approximate $\bar{v}\sim v_0$  and identify $T_{eff}=v_0^2 D_r / J \mu^2 k$.

\subsection*{Finite-size effects and flocking transition}
\begin{figure}[!h]
\centering
\includegraphics[width=.475\textwidth]{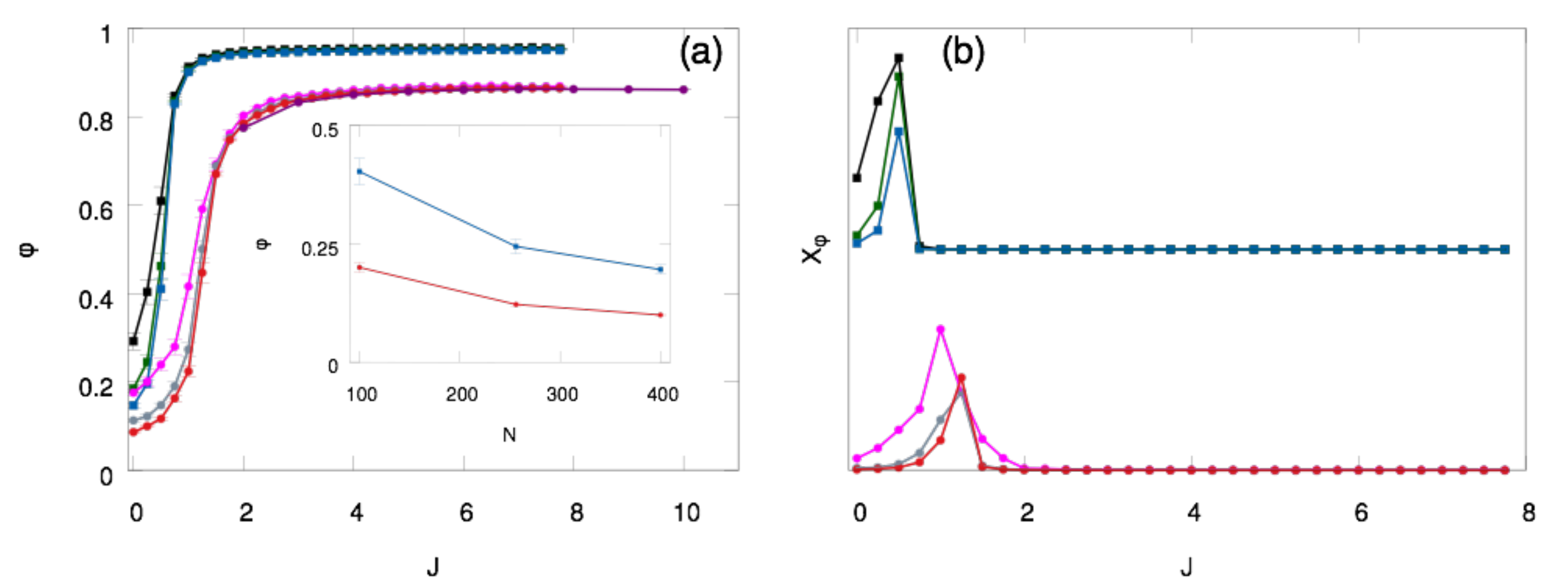}
\caption{Flocking transition. (a) Order parameter $\varphi$ 
as a function of $J$ in the solid phase ($p_0=3.1$, squares) for $N=100,256,400$ 
(black, green, and blue symbols) and in the liquid phase ($p_0=3.7$, circles) 
for $N=100,256,400,900$ (pink, gray, red, and purple). For small $J$ the
polarization $\varphi$ goes rapidly to zero with increasing system size.
In the flocking phase $\varphi=\varphi_{max}$ and the plateau value does not depend
on the system size. 
Inset: $\varphi$ for $J=0.25$ as a function of $N$.
(b) Suceptibility $\chi_{\varphi}$ as a function of $J$ for different system sizes
in the solid (top) and liquid (bottom) phases. We identify the flocking transition as the $J$ value where $\chi_\varphi$ develops a peak.}
\label{fig:sup1}      
\end{figure}

To evaluate finite-size effects on the emergence of collective
migration, we have performed numerical simulations for  system sizes 
$N=100,256,400,900$.   The order parameter $\varphi$ shown in Fig. \ref{fig:sup1}
follows a sigmoidal curve as a function of $J$. The curve is bounded between a residual polarization $\varphi_{min}(N)$ at small $J$ that decreases with increasing system size and is expected to vanish as $N\rightarrow\infty$ (inset in \ref{fig:sup1}), and an upper value
$\varphi_{max}$ at large $J$.
The transition point $J_c$ has been evaluated through the susceptibility $\chi_\varphi=N\langle (\varphi - \langle \varphi\rangle)^2\rangle$ that develops a peak at intermediates $J$ between the two plateau $\varphi_{min}$ and $\varphi_{max}$.
In the solid  the peak occurs at
$J_c^{solid}=D_r$. In the liquid the transition takes place at $J_c^{liquid}>J_c^{solid}$. 
Moreover, the plateau values $\varphi_{max}^{liquid,solid}$ do not depend on the system size. In the liquid
we obtain systematically $\varphi_{max}^{liquid}<\varphi_{max}^{solid}$ with $\varphi_{max}^{solid}\to1$
and $\varphi_{max}^{liquid}\sim 0.8$.

\subsection*{Neighbors mean-square separation}
The estimate of the transition curve $J_{flock}(p_0)$ presented in the main text is based on the numerical computation of the cage lifetime $\tau_{cage}$ associated with the neighbors mean-square separation $MSS_{nn}(t)$, evaluated for $J=0$.
The neighbors mean-square separation is defined as:
\BEQ\label{msdnn}
MSS_{nn}(t)=\frac{1}{2N}\langle \sum_{i}\sum_{j(i)}[ {\mathbf{r}_i(t)-\mathbf{r}_j(t)}]^2\rangle\;,
\EEQ 
where the sum is performed over all cells $i$ and over the two cells $j(i)$ that, at time $t=0$, are the third and the fourth nearest neighbors of cell $i$, respectively.
This choice is motivated by the following observation. Each cell $i$ is on average in contact  with six neighbors that constitute its "cage". Let $d_{cage}$ be the average distance between cell $i$ and its neighbors. 
At a given time, the first three nearest neighbors are typically closer to cell $i$ than $d_{cage}$ and thus their distance to cell $i$ will tend to increase, at least for short times. On the contrary, higher order neighbors on average will move toward cell $i$. The third and the fourth nearest neighbors are those for which these systematic effects are expected to be less important.
For example, in the solid phase, with our definition we find $MSS_{nn}(t)\approx \text{const}$, for all $t$. A different definition, based for example on the first and the second nearest neighbors, would, on the contrary, show an increase at short times, followed by damped oscillation about an asymptotic value.

Our definition has the advantage of providing an indicator that is relatively insensitive to the intra-cage dynamics, allowing us to identify unambiguously the moment when two cells, initially in contact,  start moving apart from each other, "breaking the cage".
Operatively, the cage timelife $\tau_{cage}$ is estimated as the time needed to double the neighbors mean-square separation with respect to its value at $t=0$,
\BEQ 
MSS_{nn}(\tau_{cage})\equiv 2MSS_{nn}(0)\;.
\EEQ

\subsection*{Correlation function of cell displacements}

To quantify the dynamical heterogeneities and highlight their anisotropic structure in the flocking state, we have evaluated the spatial correlation of cell displacements $C(x_\parallel,x_\perp)$ along directions longitudinal ($x_\parallel$) and transverse ($x_\perp$) to that of mean motion.

We first calculate a coarse-grained map $\Delta \rr(x,y)$ of cell displacements during a time interval $\tau_{\alpha}=10^2$ on a lattice $(x,y)$ of linear size $\delta_{\ell}=\sqrt{A_0}$.
For a given realization we calculate the spatial correlation,
\BEQ 
c(x,y)= \frac{\sum_{x^\prime, y^\prime} \Delta \rr (x+x^\prime,y+y^\prime) \cdot \Delta \rr(x^\prime,y^\prime)}{\sum_{x^\prime,y^\prime} |\Delta \rr(x^\prime,y^\prime)|^2} \; .
\EEQ 
$C(x_\parallel,x_\perp)$ is obtained by averaging $c$ over $10^2$ independent realizations. The average is performed after rotating the axes by the angle $\theta$ identifying the average direction of migration in each sample:,
\BEQ 
C(x_\parallel,x_\perp) = \left\langle  c(x_\parallel \cos \theta- x_\perp \sin \theta ,x_\perp \cos \theta+x_\parallel \sin \theta)  \right\rangle \;.
\EEQ

\subsection*{Overlap and non-Gaussian parameter}
\begin{figure}[!t]
\centering
\includegraphics[width=.475\textwidth]{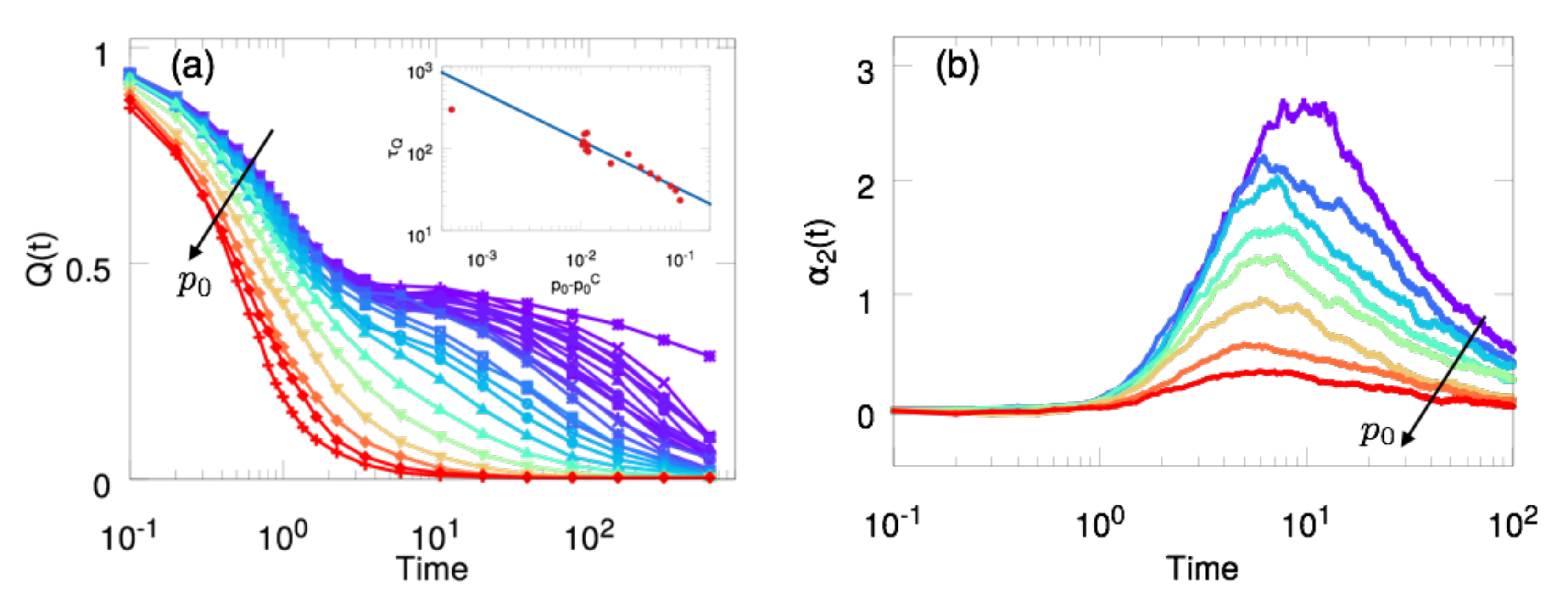}
\caption{Dynamical arrest and flocking transition.
(a) The overlap $Q(t)$ for $J=2.0$ shows a clear two step decay typical of glassy dynamics. From $Q(t)$ we can extract a correlation time $\tau_q$ (inset) that increases as a power law as a function of $p_0-p_0^C$.
(b) The non-Gaussian parameter $\alpha_2(t)$ develops a clear peak indicating the presence 
of angular dynamical heterogeneity in the flocking state.	}
\label{fig:msd}
\end{figure}

To estimate the effect of alignment on structural rearrangements we have also looked at the behavior of the 
overlap parameter, $Q(t)$. The overlap gives a measure of the similarity between two configurations of the system 
taken at two different times, in our case at $t$ and $0$ \cite{Berthier11,KobGlotzer1997}.
To compute $Q(t)$, we discretize space in a lattice of linear size $\delta \sim 0.66$ and define $n_i(t)=1$ if the site $i$ s occupied by the same particle at time $0$ and $t$, and $n_i(t)=0$ otherwise. 
The overlap is defined as 
\BEQ 
Q(t)=\frac{ \sum_i n_i(t) n_i(0) }{\sum_{i} n_i(0)} \; .
\EEQ 
In order to exclude fast vibrations on short time scales,
we choose the parameter $\delta$ through the condition $\delta=\sqrt{MSD(t_{s-d})}$, where $t_{s-d}$ is the crossover time 
from subdiffusive to diffusive regime.
As one can appreciate looking at Fig. \ref{fig:msd}(a), where Q(t) for $J=2.0$ is shown, 
the overlap undergoes a two-step decay typical of glassy systems indicating a crossover between
fast and slow processes. Remarkably, the crossover takes place around the flocking transition
indicating that the emergence of collective migration changes the structural properties. This is also highlight in the main  text through the study of dynamical heterogeneities.

In the inset of the same figure, we plot $\tau_Q \equiv \int_{t^*}^\infty \, dt\,Q(t)$, where $t^*$ is chosen in a way to extract only the slow $\beta$ decay of $Q(t)$. Interestingly, $\tau_Q$ can be fitted to a power law as a function of $p_0-p_0^C$, which is compatible with mode-coupling theory \cite{Berthier11}.

The time scale $\tau_{Q}$ is also compatible with
the time scale of the peak in the non-Gaussian parameter of the angular
displacements $\alpha_2(t)\equiv \frac{1}{3}\frac{\langle \Delta \theta^4 \rangle}{\langle \Delta \theta^2 \rangle^2} -1$, 
where $\Delta \theta^a \equiv \sum_i N^{-1} \left[ \theta_i (t) - \theta_i(0) \right]^a$. The behavior of $\alpha_2(t)$ is shown in  Fig.~\ref{fig:msd}(b).

\bibliography{bibflock}

\end{document}